\documentclass[11pt]{article}
\usepackage{moriond,epsfig}
\usepackage{subfigure}

\def\be{\begin{equation}}
\def\ee{\end{equation}}
\def\bea{\begin{eqnarray}}
\def\eea{\end{eqnarray}}

\def\Etmiss{\hbox{$\rlap{\kern0.25em/}E_T$}}

\begin{document}
\vspace*{4cm}
\title{MODEL INDEPENDENT SEARCHES FOR NEW PHYSICS AT THE TEVATRON}

\author{J. PIPER}

\address{Department of Physics and Astronomy, Michigan State University,\\ 
Biomedical and Physical Sciences Building,\\
East Lansing, MI, USA 48824\\
\vspace{5mm}
for the CDF and D\O\ collaborations}

\maketitle\abstracts{The standard model is a successful but limited theory. There is significant theoretical motivation to believe that new physics may appear at the energy scale of a few TeV, the lower end of which is currently probed by the Fermilab Tevatron Collider. The methods used to search for physics beyond the standard model in a model independent way and the results of these searches based on 1.0 fb$^{-1}$ of data collected with the D\O\ detector and 2.0 fb$^{-1}$ at the CDF detector are presented.}

\section{Introduction and Strategies}
 Particle physics is at a stage where there is no
unique way forward. The standard model of particle physics
has been remarkably successful: all fundamental particles predicted by this
model have been discovered, with the notable exception of the Higgs boson.
Despite its success, there are strong motivations from the theory
to expect new physics at energies at or just above the electroweak
scale.

Assuming that new physics does exist, we do not know what this new
physics is and thus precisely how to search for it. There are many
theories and models which predict differences between reality and
the standard model. But generally these theories and models do not give precise
energy and phase space regions to search for new physics. Motivated by this, a scan over many channels was performed at each experiment to look for significant deviations from 
the standard model. The CDF experiment searches over all high-$p_T$ data. At D0, it was found that the background model was most developed for the case of final states
containing leptons. Similar approaches to search for new physics have 
been applied to data from D\O\ Run I,~\cite{run1sleuth1,run1sleuth2,run1sleuth3} and
H1 at HERA~\cite{heraMIS}.


At CDF and D\O\ , several methods are used to search across large portions of the data set. First, exclusive final states are tested for agreement in event counts using the {\sc vista} algorithm. {\sc vista} also searches many histograms for each of the final states for disagreement. In {\sc sleuth}, one variable, $(\Sigma p_{T,obj}) + MET$, is used for the search, focusing on the tails of the distributions. Finally, at CDF, masses of all object combinations are searched for discrepancies in {\sc bump hunter}.

CDF \cite{cdfPRD,cdfRC} focuses on consistency and reach. The objects used in all final states have the same set of object cuts. The standard model background implementation comes from Monte Carlo, primarily {\sc pythia}\cite{pythia} and {\sc MadEvent}\cite{madevent}. After the Monte Carlo is pushed through the detector simulation, additional corrections to fix modeling issues are performed in a standard, rigorous way. All of the final states and histograms are put into one immense fit, using as few parameters as possible to provide basic agreement across final states. The fit uses all of the exclusive final states defined in {\sc vista} with the histograms provided for those final states. In practice, it is often one histogram that dominates the value found for any of the individual parameters. CDF introduced 43 correction parameters to minimize final state discrepancies. The parameters are further constrained by adding information external to that used in the {\sc vista} high-$p_T$ data sample. The correction factors obtained from the fit are found to be consistent with those derived in control regions of other analyses at CDF.

As at CDF, D\O\ runs Monte Carlo simulations to predict the standard model expectation. D\O\ uses mostly {\sc alpgen}\cite{alpgen} and {\sc pythia} Monte Carlo. D\O\ models multijet background from data rather than Monte Carlo.  To estimate these backgrounds, some of the object selection cuts are reversed in data to produce samples with electron, 
muon, or tau objects which are mostly misidentified jets from multijet backgrounds.

D\O\ then divides the whole data set into seven nonoverlapping final states. In each of these, previously determined standard weights are applied using well-understood areas of phase space dominated by particular standard model processes to account for necessary corrections to the Monte Carlo and detector modeling. These seven states are defined by their lepton content and are inclusive in jets and additional objects. A fit is then performed for each of these states 
to obtain the scale factors which reproduce the distributions of the 
selected data with the background from Monte Carlo and multijet background determined from data.

\section{General Searches Using {\sc vista}}
 
{\sc vista} performs two checks,
a normalization-only check on the number of events in each
exclusive state, and a
shape-only analysis of histograms within a state by calculating a
Kolmogorov-Smirnov statistic (and resulting fit probability).
Both of these numbers require additional interpretation, because
of the number of trials involved. When observing many final states, some disagreements are expected merely due to statistical fluctuations in the data.  Therefore, probabilities calculated with both of these methods are corrected to reflect this multiple testing.

\begin{figure}
\begin{center}
\subfigure[] { 
\includegraphics[height=1.5in]{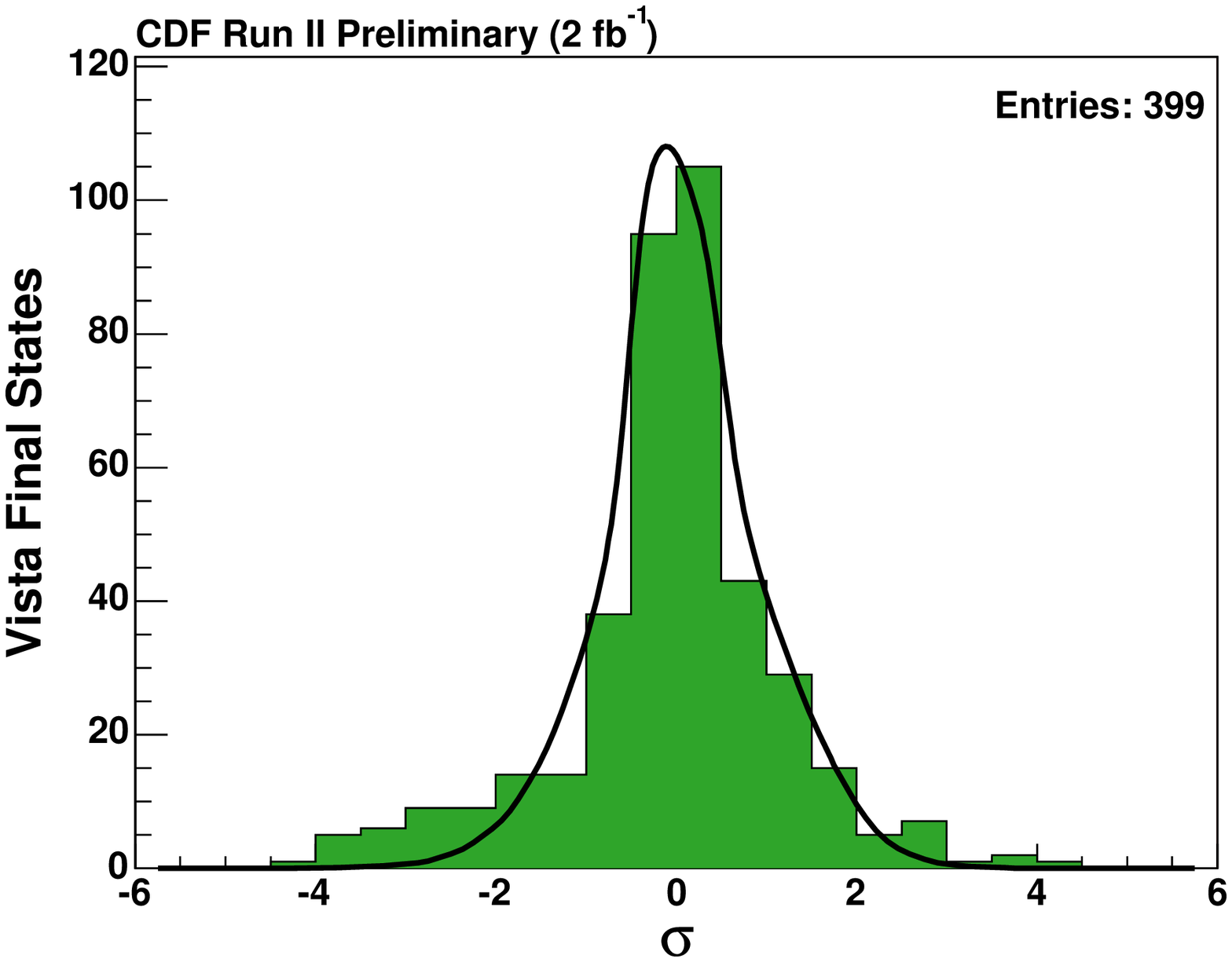}
\label{fig:cdf_fs}
}
\subfigure[] { 
\includegraphics[height=1.5in]{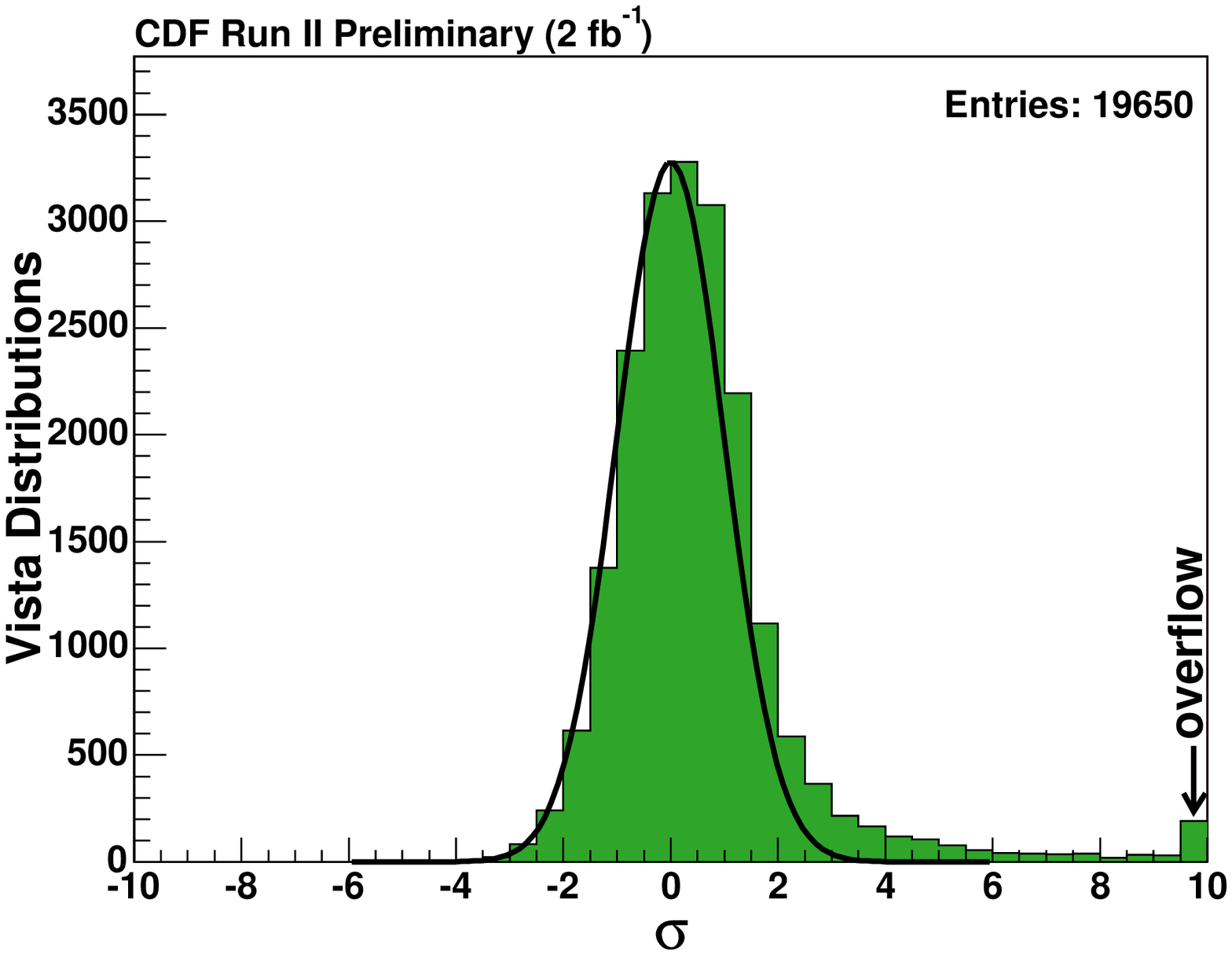}
\label{fig:cdf_shape}
}
\subfigure[] { 
\includegraphics[height=1.5in]{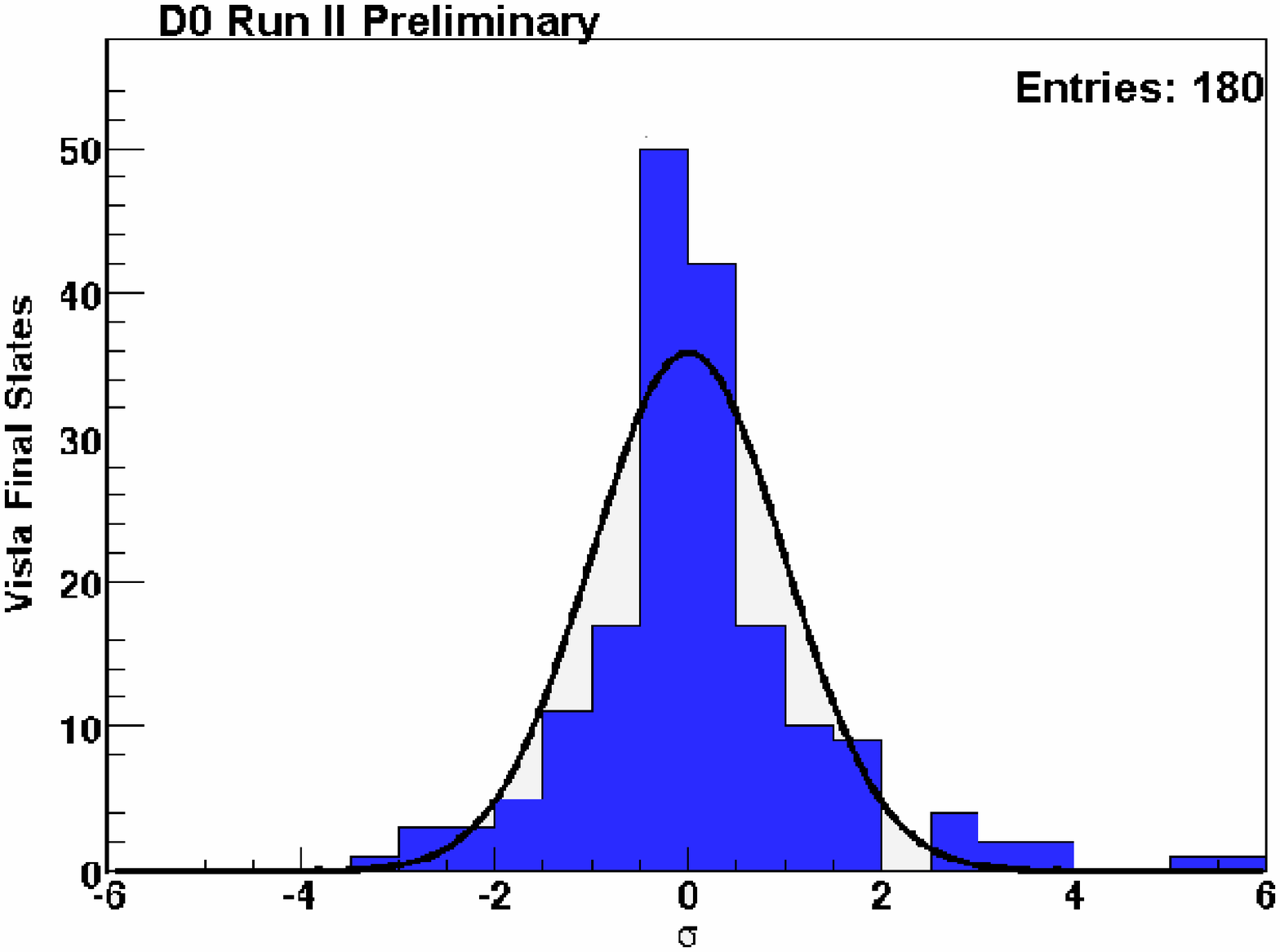}
\label{fig:d0_fs}
}
\subfigure[] { 
\includegraphics[height=1.5in]{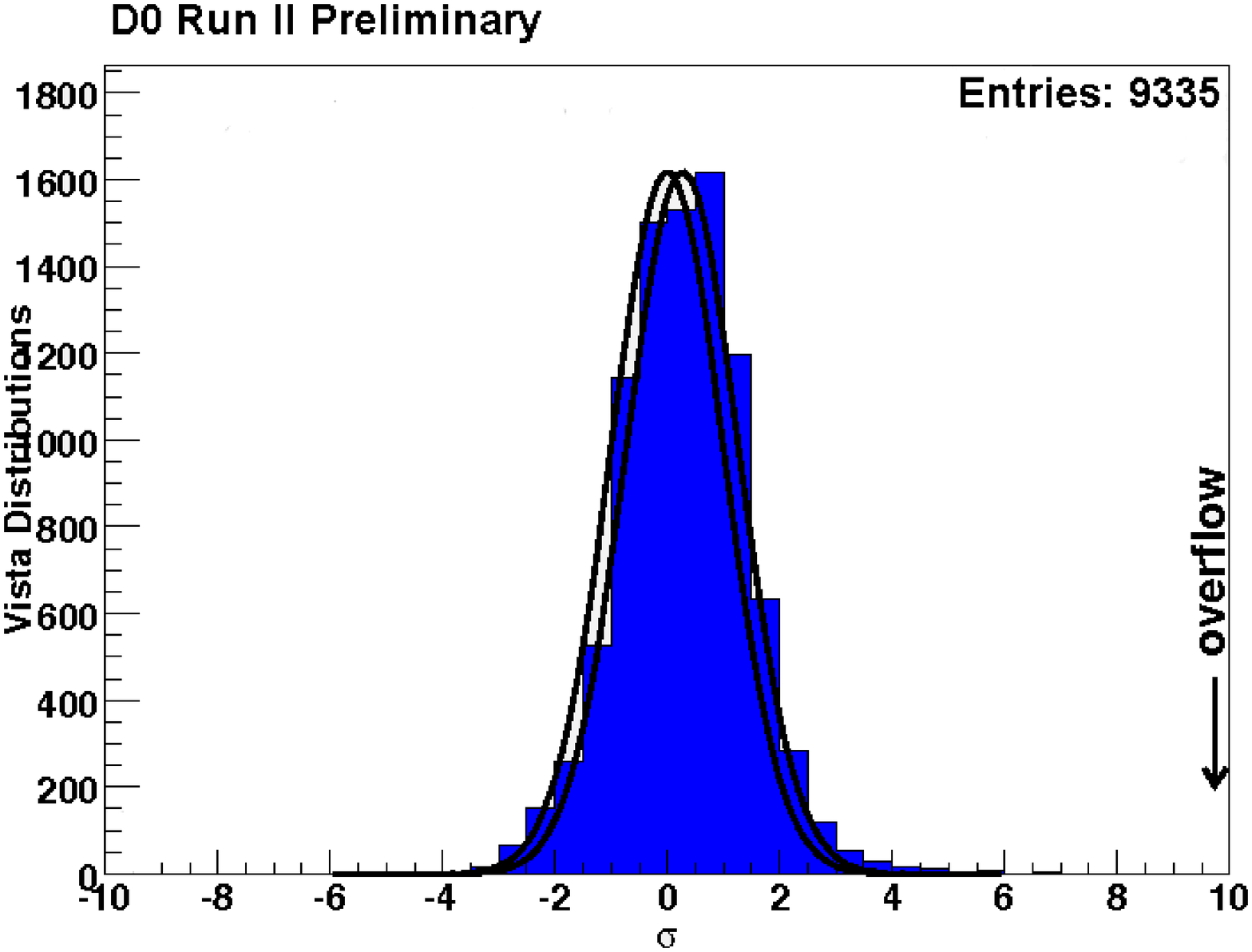}
\label{fig:d0_shape}
}
\caption{CDF and D\O\ distributions of final state counts and Kolmogorov-Smirnov probabilities converted to units of $\sigma$ before trials factor. The CDF final state and shape distributions are shown in Figs. \ref{fig:cdf_fs} and \ref{fig:cdf_shape} while the corresponding D\O\ distributions are shown in Figs. \ref{fig:d0_fs} and \ref{fig:d0_shape}.}
\end{center}
\end{figure}

At CDF, there are no discrepancies in event counts over the 399 states checked. Additionally, 555 1-D shape histograms show discrepancy after trials. Many of these can be attributed to a Monte Carlo ``3-jet effect''. The ``3-jet effect'' is seen in several types of histograms across many of the multijet final states. These include the $\Delta \mathcal{R} (j_2,j_3)$ and the jet mass distributions. Examining different parton shower generators show that predictions for this region of phase space are not consistent. These predictions have not yet been compared against LEP I data to provide reasonable constraints on this effect. The distributions of final state counts and histograms are shown is Figs. \ref{fig:cdf_fs} and \ref{fig:cdf_shape}.

At D\O\ , of the 180 distributions, four show
significant event count discrepancies. These are the final states $\mu$ + 2 jets + $\Etmiss$ 
with a converted probability of 9.3$\sigma$ after trials correction, $\mu$ + $\gamma$ + 1 jet + $\Etmiss$ with 6.6$\sigma$, $\mu^{+} \mu^{-}$ + $\Etmiss$ with a discrepancy of 4.4$\sigma$ and $\mu^{+} \mu^{-}$ + $\gamma$ at 4.1$\sigma$. Additionally, 24 histograms show shape discrepancies. The distributions of final state counts and shape discrepancies can be seen in Figs. \ref{fig:d0_fs} and \ref{fig:d0_shape}.

Two of these states are directly related to oversimplified modeling of the photon misidentification rate. The $\mu$ + 2 jets + $\Etmiss$ final state discrepancy shows an excess of events with a muon at $\eta > 1.0$. The excess points to an oversimplification in our approach to trigger efficiencies. The proportion of events that are brought in by single muon vs. muon plus jets triggers changes significantly as we increase jet multiplicity. These triggers introduce $\eta$-dependent efficiencies which are not fully incorporated into our simple fits. The dimuon with missing energy final state shows an excess of data compared to the standard model Monte Carlo prediction. A study into the track curvature of data and MC muons, and of the associated resolution, has shown that an additional smearing should be applied in the Monte Carlo to appropriately simulate very high $p_T$ muons. The prime signature of these muons is an excess of $\Etmiss$ because of the lack of compensation for the mismeasured, unbalanced track.

\section{Targeted Searches using {\sc bump hunter} (CDF) and {\sc sleuth}}

The {\sc bump hunter} algorithm scans the spectrum of most mass variables with a sliding window. The window varies across each final state and within final states themselves to account for changing detector resolution. For each combination of objects, the mass is calculated in each sliding window, and the probability to see a data excess as large as the one observed is noted. The window then slides by two times the mass resolution. In order for a bump to be defined, the region must contain at least five data events, and the side bands (regions on either side of the region of interest and of half the width) must be less discrepant than the central region of interest. The smallest probability for each final state is determined and the overall significance of the most significant of these excesses is evaluated by performing pseudoexperiments. 

One bump is found to cross the discovery threshold. This is the mass of four jets in a final state with four jets of $\sum p_T~<~400$. This discrepancy is attributed to the same three jet effect as was observed previously in the {\sc vista} shape discrepancies. The same effect is observed in other final states of different jet multiplicities with a significance that falls below the threshold.

In {\sc sleuth}, $\sum p_T$ in each channel is searched for a
cut which maximizes the significance of the data excess over the SM
backgrounds producing $\sum p_T > \mathrm{cut}$.  This
significance is then corrected for the number of trials in both
the number of possible positions of the cut in the histogram, and
at a higher level, the number of final states Sleuth examines. The final corrected probability corresponds to the
probability that an individual final state would produce one or more
probabilities as small as observed.  A significant output from Sleuth is defined as a state with corrected
probability $<~0.001$ (equivalent to about 3 Gaussian standard deviations).  

CDF finds no discrepant state that crosses the discovery threshold. The most discrepant states at CDF contain same sign muon-electron pairs. The most significant of these causes a $\tilde{\mathcal{P}}$ of 0.085 which is still well above the threshold of 0.001. Sleuth by definition does not focus on the correlation among the most discrepant objects and instead focuses on just the single most discrepant state.

The {\sc vista} final states that show broad numerical excesses at D\O\ are found again with the {\sc sleuth} algorithm, as would be expected. One additional distribution crosses the discovery threshold of $\tilde{\mathcal{P}}$ \textless~0.001, where $\tilde{\mathcal{P}}$ is the probability after all trial factors: the final state that crosses the discovery threshold is $\mu^{\pm}$ + $e^{\mp}$ + $\Etmiss$. Currently the evidence suggests that the muon tracking resolution is responsible for this discrepancy. A large fraction of the events in the tail of the Sleuth distribution have a muon with a very large $p_T$ and large missing energy. With the present modeling of muon resolution, straight track events are underrepresented in the standard model background estimation. This state has 46 data events in the tail of this distribution compared to only 17 predicted by the Monte Carlo.

\section{Conclusion}

In conclusion, CDF and D\O\ have performed broad searches for new physics using $2.0~fb^{-1}$ at CDF and $1.0~fb^{-1}$ at D\O\ . Hundreds of exclusive data final states and thousands of kinematic distributions were compared to the 
complete standard model background predictions at each experiment using the {\sc vista} algorithm. No final states were discrepant at CDF and only four out of 180 exclusive final states showed a statistically 
significant discrepancy at D\O\ . Given the known modeling difficulties in all four final states, we refrain from attributing the observed discrepancies to new physics. 
A quasi-model independent search for new physics was also 
performed using the algorithm Sleuth by looking at the regions of 
excess on the high-$\sum p_T$ tails of exclusive final states.
Only $\mu^{\pm}$ + $e^{\mp}$ + $\Etmiss$ 
surpasses the discovery threshold beyond the obvious excesses noticed in {\sc vista}, and this seems to be related to difficulties in modeling the muon $p_T$ resolution. At CDF, there were no Sleuth discrepancies, but the most discrepant final states contained a same-sign muon and electron. Since no final state met the criteria defined beforehand for discovery, no new physics claim is made. Although there were no strong hints of new physics in the data, a factor of five more data has already been recorded at each experiment. As this data is incorporated into the analyses and improvements are implemented in the correction models, both experiments should be much more sensitive to possible new physics. 

\end{document}